\documentclass[twoside]{article}
\usepackage{amsfonts}
\usepackage{amssymb}

\catcode`\@=11 \long\def\@makefntext#1{ \protect\noindent \hbox
to 3.2pt {\hskip-.9pt
$^{{\eightrm\@thefnmark}}$\hfil}#1\hfill}       

\def\thefootnote{\fnsymbol{footnote}}
\def\@makefnmark{\hbox to 0pt{$^{\@thefnmark}$\hss}}    

\def\ps@myheadings{\let\@mkboth\@gobbletwo
\def\@oddhead{\hbox{}
\rightmark\hfil\eightrm\thepage}
\def\@oddfoot{}\def\@evenhead{\eightrm\thepage\hfil
\leftmark\hbox{}}\def\@evenfoot{}
\def\sectionmark##1{}\def\subsectionmark##1{}}

\oddsidemargin=\evensidemargin 
\renewcommand{\thefootnote}{\fnsymbol{footnote}}

\newcounter{sectionc}\newcounter{subsectionc}\newcounter{subsubsectionc}
\renewcommand{\section}[1] {\vspace{12pt}\addtocounter{sectionc}{1}
\setcounter{subsectionc}{0}\setcounter{subsubsectionc}{0}\noindent
    {\tenbf\thesectionc. #1}\par\vspace{5pt}}
\renewcommand{\subsection}[1] {\vspace{12pt}\addtocounter{subsectionc}{1}
    \setcounter{subsubsectionc}{0}\noindent
    {\bf\thesectionc.\thesubsectionc. {\kern1pt \bfit #1}}\par\vspace{5pt}}
\renewcommand{\subsubsection}[1] {\vspace{12pt}\addtocounter{subsubsectionc}{1}
    \noindent{\tenrm\thesectionc.\thesubsectionc.\thesubsubsectionc.
    {\kern1pt \tenit #1}}\par\vspace{5pt}}
\newcommand{\nonumsection}[1] {\vspace{12pt}\noindent{\tenbf #1}
    \par\vspace{5pt}}

\newcounter{appendixc}
\newcounter{subappendixc}[appendixc]
\newcounter{subsubappendixc}[subappendixc]
\renewcommand{\thesubappendixc}{\Alph{appendixc}.\arabic{subappendixc}}
\renewcommand{\thesubsubappendixc}
    {\Alph{appendixc}.\arabic{subappendixc}.\arabic{subsubappendixc}}

\renewcommand{\appendix}[1] {\vspace{12pt}
        \refstepcounter{appendixc}
        \setcounter{figure}{0}
        \setcounter{table}{0}
        \setcounter{lemma}{0}
        \setcounter{theorem}{0}
        \setcounter{corollary}{0}
        \setcounter{definition}{0}
        \setcounter{equation}{0}
        \renewcommand{\thefigure}{\Alph{appendixc}.\arabic{figure}}
        \renewcommand{\thetable}{\Alph{appendixc}.\arabic{table}}
        \renewcommand{\theappendixc}{\Alph{appendixc}}
        \renewcommand{\thelemma}{\Alph{appendixc}.\arabic{lemma}}
        \renewcommand{\thetheorem}{\Alph{appendixc}.\arabic{theorem}}
        \renewcommand{\thedefinition}{\Alph{appendixc}.\arabic{definition}}
        \renewcommand{\thecorollary}{\Alph{appendixc}.\arabic{corollary}}
        \renewcommand{\theequation}{\Alph{appendixc}.\arabic{equation}}
        \noindent{\tenbf Appendix \theappendixc #1}\par\vspace{5pt}}
\newcommand{\subappendix}[1] {\vspace{12pt}
        \refstepcounter{subappendixc}
        \noindent{\bf Appendix \thesubappendixc. {\kern1pt \bfit #1}}
    \par\vspace{5pt}}
\newcommand{\subsubappendix}[1] {\vspace{12pt}
        \refstepcounter{subsubappendixc}
        \noindent{\rm Appendix \thesubsubappendixc. {\kern1pt \tenit #1}}
    \par\vspace{5pt}}

\newcommand{\textlineskip}{\baselineskip=13pt}
\newcommand{\smalllineskip}{\baselineskip=10pt}

\def\eightcirc{
\begin{picture}(0,0)
\put(4.4,1.8){\circle{6.5}}
\end{picture}}
\def\eightcopyright{\eightcirc\kern2.7pt\hbox{\eightrm c}}

\def\abstracts#1#2#3{{
    \centering{\begin{minipage}{4.5in}\baselineskip=10pt\footnotesize
    \parindent=0pt #1\par
    \parindent=15pt #2\par
    \parindent=15pt #3
    \end{minipage}}\par}}

\renewenvironment{thebibliography}[1]
    {\frenchspacing
     \ninerm\baselineskip=11pt
     \begin{list}{\arabic{enumi}.}
    {\usecounter{enumi}\setlength{\parsep}{0pt}
     \setlength{\leftmargin 12.7pt}{\rightmargin 0pt} 
     \setlength{\itemsep}{0pt} \settowidth
    {\labelwidth}{#1.}\sloppy}}{\end{list}}

\newcounter{itemlistc}
\newcounter{romanlistc}
\newcounter{alphlistc}
\newcounter{arabiclistc}

\newcommand{\fcaption}[1]{
        \refstepcounter{figure}
        \setbox\@tempboxa = \hbox{\footnotesize Fig.~\thefigure. #1}
        \ifdim \wd\@tempboxa > 5in
           {\begin{center}
        \parbox{5in}{\footnotesize\smalllineskip Fig.~\thefigure. #1}
            \end{center}}
        \else
             {\begin{center}
             {\footnotesize Fig.~\thefigure. #1}
              \end{center}}
        \fi}

\newcommand{\tcaption}[1]{
        \refstepcounter{table}
        \setbox\@tempboxa = \hbox{\footnotesize Table~\thetable. #1}
        \ifdim \wd\@tempboxa > 5in
           {\begin{center}
        \parbox{5in}{\footnotesize\smalllineskip Table~\thetable. #1}
            \end{center}}
        \else
             {\begin{center}
             {\footnotesize Table~\thetable. #1}
              \end{center}}
        \fi}

\def\@citex[#1]#2{\if@filesw\immediate\write\@auxout
    {\string\citation{#2}}\fi
\def\@citea{}\@cite{\@for\@citeb:=#2\do
    {\@citea\def\@citea{,}\@ifundefined
    {b@\@citeb}{{\bf ?}\@warning
    {Citation `\@citeb' on page \thepage \space undefined}}
    {\csname b@\@citeb\endcsname}}}{#1}}

\newif\if@cghi
\def\cite{\@cghitrue\@ifnextchar [{\@tempswatrue
    \@citex}{\@tempswafalse\@citex[]}}
\def\citelow{\@cghifalse\@ifnextchar [{\@tempswatrue
    \@citex}{\@tempswafalse\@citex[]}}
\def\@cite#1#2{{$\null^{#1}$\if@tempswa\typeout
    {IJCGA warning: optional citation argument
    ignored: `#2'} \fi}}

\def\pmb#1{\setbox0=\hbox{#1}
    \kern-.025em\copy0\kern-\wd0
    \kern.05em\copy0\kern-\wd0
    \kern-.025em\raise.0433em\box0}

\def\fnm#1{$^{\mbox{\scriptsize #1}}$}
\def\fnt#1#2{\footnotetext{\kern-.3em
    {$^{\mbox{\scriptsize #1}}$}{#2}}}

\def\fpage#1{\begingroup
\voffset=.3in
\thispagestyle{empty}\begin{table}[b]\centerline{\footnotesize #1}
    \end{table}\endgroup}

\def\runninghead#1#2{\pagestyle{myheadings}
\markboth{{\protect\footnotesize\it{\quad #1}}\hfill}
{\hfill{\protect\footnotesize\it{#2\quad}}}} \headsep=15pt

\font\tenrm=cmr10 \font\tenit=cmti10 \font\tenbf=cmbx10
\font\bfit=cmbxti10 at 10pt \font\ninerm=cmr9 
 \font\eightrm=cmr8

\def\qed{\hbox{${\vcenter{\vbox{            
   \hrule height 0.4pt\hbox{\vrule width 0.4pt height 6pt
   \kern5pt\vrule width 0.4pt}\hrule height 0.4pt}}}$}}

\renewcommand{\thefootnote}{\fnsymbol{footnote}}    
\renewcommand{\theequation}{\thesectionc.\arabic{equation}}
\csname @addtoreset\endcsname{equation}{section}

\begin{document}

\runninghead{Representations of superconformal algebras} {in the AdS${}_{7/4}$/CFT${}_{6/3}$ correspondence}

\normalsize\textlineskip \thispagestyle{empty}
\setcounter{page}{1}

\begin{flushright}
CERN-TH/2000-298
\\ LAPTH-815/2000
\end{flushright}
\vspace{.5cm} \vspace*{0.88truein}

\fpage{1} \centerline{\bf Representations of superconformal algebras}
\vspace*{0.015truein}
\centerline{\bf in the AdS${}_{7/4}$/CFT${}_{6/3}$ correspondence }
\vspace*{0.37truein} \centerline{\footnotesize Sergio Ferrara}
\vspace*{0.015truein} \centerline{\footnotesize\it CERN
Theoretical Division } \baselineskip=10pt
\centerline{\footnotesize\it CH 1211 Geneva 23, Switzerland}
\baselineskip=10pt
\centerline{\footnotesize\it and Laboratori Nazionali di Frsacati, INFN, Italy}
\vspace*{10pt} \centerline{\footnotesize Emery Sokatchev}
\vspace*{0.015truein} \centerline{\footnotesize\it Laboratoire
d'Annecy-le-Vieux de Physique Th\'{e}orique LAPTH\footnote{UMR 5108
associ{\'e}e {\`a} l'Universit{\'e} de Savoie} - BP 110 }
\baselineskip=10pt \centerline{\footnotesize\it F-74941
Annecy-le-Vieux Cedex, France} \vspace*{0.225truein}

\vspace*{0.21truein} \abstracts{We perform a general analysis of representations of the superconformal algebras $\mbox{OSp}(8/4,\mathbb{R})$ and
$\mbox{OSp}(8^*/2N)$ in harmonic superspace. We present a
construction of their highest-weight UIR's by multiplication of
the different types of massless conformal superfields
(``supersingletons").
\\ In particular, all ``short multiplets" are classified.
Representations undergoing shortening have ``protected dimension"
and may correspond to BPS states in the dual supergravity theory
in anti-de Sitter space.\\ These results are relevant for the
classification of multitrace operators in boundary conformally
invariant theories as well as for the classification of AdS black
holes preserving different fractions of supersymmetry.}{}{}


\vspace*{1pt}\textlineskip  
\setcounter{footnote}{0}
\renewcommand{\thefootnote}{\alph{footnote}}

\section{Introduction}    
\vspace*{-0.5pt}
\noindent

Superconformal algebras and their representations play a crucial
r\^{o}le in the AdS/CFT correspondence because of their dual r\^{o}le of
describing the gauge symmetries of the AdS bulk supergravity
theory and the global symmetries of the boundary conformal field
theory \cite{mal,gkp,wit}.

A special class of configurations which are particularly relevant
are the so-called BPS states, i.e. dynamical objects corresponding
to representations which undergo ``shortening".

These representations can only occur when the conformal dimension
of a (super)primary operator is ``quantized" in terms of the R
symmetry quantum numbers and they are at the basis of the
so-called ``non-renormalization" theorems of supersymmetric
quantum theories \cite{FIZ}.

There exist different methods of classifying the UIR's of
superconformal algebras. One is the so-called oscillator
construction of the Hilbert space in which a given UIR
acts \cite{bgg}. Another one, more appropriate to
describe field theories, is the realization of such
representations on superfields defined in superspaces
\cite{SS,fwz}. The latter are ``supermanifolds" which can be
regarded as the quotient of the conformal supergroup by some of
its subgroups.

In the case of ordinary superspace the subgroup in question is the
supergroup obtained by exponentiating a non-semisimple
superalgebra which is the semidirect product of a super-Poincar\'{e}
graded Lie algebra with dilatation ($\mbox{SO}(1,1)$) and the R
symmetry algebra. This is the superspace appropriate for non-BPS
states. Such states correspond to bulk massive states which can
have ``continuous spectrum" of the AdS mass (or, equivalently, of
the conformal dimension of the primary fields).

BPS states are naturally associated to superspaces with lower
number of ``odd" coordinates and, in most cases, with some
internal coordinates of a coset space $G/H$. Here $G$ is the R
symmetry group of the superconformal algebra, i.e. the subalgebra
of the even part which commutes with the conformal algebra of
space-time and $H$ is some subgroup of $G$ having the same rank as
$G$.

Such superspaces are called ``harmonic" \cite{GIK1} and they are
characterized by having a subset of the initial odd coordinates
$\theta$. The complementary number of odd variables determines the
fraction of supersymmetry preserved by the BPS state. If a BPS
state preserves $K$ supersymmetries then the $\theta$'s of the
associated harmonic superspace will transform under some UIR of
$H_K$.

For 1/2 BPS states, i.e. states with maximal supersymmetry, the
superspace involves the minimal number of odd coordinates (half of
the original one) and $H_K$ is then a maximal subgroup of $G$. On
the other hand, for states with the minimal fraction of
supersymmetry $H_K$ reduces to the ``maximal torus" whose Lie
algebra is the Cartan subalgebra of $G$.

It is the aim of the present paper to give a comprehensive
treatment of BPS states related to ``short representations" of
superconformal algebras for the cases which are most relevant in
the context of the AdS/CFT correspondence, i.e. the $d=3$ ($N=8$)
and $d=6$ ($N=(2,0)$). The underlying conformal field theories
correspond to world-volume theories of $N_c$ copies of $M_2$,
$M_5$ and $D_3$ branes in the large $N_c$ limit
\cite{AOY}-\cite{ckvp} which are ``dual" to AdS supergravities
describing the horizon geometry of the branes \cite{AGMOO}.

The present contribution summarizes the results which have
already appeared elsewhere \cite{FS2,FS3,FS4}. We first carry out
an abstract analysis of the conditions for Grassmann
(G-)analyticity \cite{GIO} (the generalization of the familiar
concept of chirality \cite{fwz}) in a superconformal context. We
find the constraints on the conformal dimension and R symmetry
quantum numbers of a superfield following from the requirement
that it do not depend on one or more Grassmann variables.
Introducing G-analyticity in a traditional superspace cannot be
done without breaking the R symmetry. The latter can be restored
by extending the superspace by harmonic variables
\cite{Rosly,GIK1,GIK11}-\cite{hh} parametrizing the coset
$G/H_K$. We also consider the massless UIR's (``supersingleton"
multiplets) \cite{ff2,Fr}, first as constrained superfields in
ordinary superspace \cite{HSiT}-\cite{Howe} and then, for a part
of them, as G-analytic harmonic superfields \cite{GIK1,hh,Howe}.
Next we use supersingleton multiplication to construct UIR's of
$\mbox{OSp}(8^*/2N)$ and $\mbox{OSp}(8/4,\mathbb{R})$. We show
that in this way one can reproduce the complete classification of
UIR's of ref. \cite{Minw2}. We also discuss different kinds of
shortening which certain superfields (not of the BPS type) may
undergo. We conclude the paper by listing the various BPS states
in the physically relevant cases of $M_2$ and $M_5$ branes horizon
geometry where only one type of supersingletons appears.

Massive towers corresponding to 1/2 BPS states are the K-K modes
coming from compactification of M-theory on $AdS_{7/4}\times
S_{4/7}$ \cite{masstow,AOY}. Short representations of
superconformal algebras also play a special r\^ole in determining
$N$-point functions from OPE \cite{dHP,corbast}.

Another area of interest is the classification of AdS black holes
\cite{hawk}-\cite{duffl}, according to the fraction of
supersymmetry preserved by the black hole background.

In a parallel analysis with black holes in asymptotically flat
background \cite{FMG}, the AdS/CFT correspondence predicts that
such BPS states should be dual to superconformal states undergoing
``shortening" of the type discussed here.

\setcounter{equation}0
\section{The six-dimensional case}

In this section we describe highest-weight UIR's of the
superconformal algebras $\mbox{OSp}(8^*/2N)$ in six dimensions.
Although the physical applications refer to $N=1$ and $N=2$, it
is worthwhile to carry out the analysis for general $N$, along
the same lines as in the four-dimensional case \cite{AFSZ,FS1}. We
first examine the consequences of G-analyticity and conformal
supersymmetry and find out the relation to BPS states. Then we
will construct UIR's of $\mbox{OSp}(8^*/2N)$ by multiplying
supersingletons. The results exactly match the general
classification of UIR's of $\mbox{OSp}(8^*/2N)$ of Ref.
\cite{Minw2}.

\subsection{The conformal superalgebra $\mbox{OSp}(8^*/2N)$
and Grassmann analyticity}

The standard realization of the conformal superalgebra
$\mbox{OSp}(8^*/2N)$ makes use of the superspace
\begin{equation}
{\mathbb R}^{6\vert 8N} = {\mbox{OSp}(8^*/2N)\over \{K,S,M,D,T\}}
= (x^\mu,\theta^{\alpha\; i}) \label{6.5}
\end{equation}
where $\theta^{\alpha\; i}$ is a left-handed spinor carrying an
index $i=1,2,\ldots,2N$ of the fundamental representation of the R
symmetry group $\mbox{USp}(2N)$. Unlike the four-dimensional case,
here chirality is not an option but is already built in. The only
way to obtain smaller superspaces is through Grassmann
analyticity. We begin by imposing a single condition of
G-analyticity on the superfield defined in (\ref{6.5}):
\begin{equation}\label{6.6}
  q^1_\alpha\Phi(x,\theta)=0
\end{equation}
which amounts to considering the coset
\begin{equation}
{\mathbb A}^{6\vert 4(2N-1)} = {\mbox{OSp}(8^*/2N)\over
\{K,S,M,D,T,Q^1\}} = (x^\mu,\theta^{\alpha\; 1,2,\ldots,2N-1})
\label{6.7}
\end{equation}
From the algebra of $\mbox{OSp}(8^*/2N)$ we obtain
\begin{eqnarray}
  &&m_{\mu\nu}=0\;, \label{6.8}\\
  &&t^{11}=t^{12}=\ldots=t^{1\; 2N-1}=0\;, \label{6.9}\\
  &&4t^{1\; 2N}+\ell=0\;.\label{6.10}
\end{eqnarray}
Eq. (\ref{6.8}) implies that the superfield $\Phi$ must be a
Lorentz scalar. In order to interpret eqs. (\ref{6.9}),
(\ref{6.10}), we need to split the generators of $\mbox{USp}(2N)$
into raising operators (corresponding to the positive roots), $
  T^{k\; 2N-l}\;, \ \ k=1,\ldots, N,\ l=k,\ldots,2N-k\quad (\mbox{simple if
$l=k$})$, $[\mbox{U}(1)]^N$ charges $
  H_k = -2 T^{k\; 2N-k+1}\;, \quad k=1,\ldots, N $ and lowering operators. The  Dynkin labels $a_k$ of a
$\mbox{USp}(2N)$ irrep are defined as follows:
\begin{equation}\label{6.113}
  a_k=H_k-H_{k+1}\;, \ \ k=1,\ldots,N-1\;, \quad a_N=H_N\;,
\end{equation}
so that, for instance, the generator $Q^1$ is the HWS of the
fundamental irrep $(1,0,\ldots,0)$.

Now it becomes clear that (\ref{6.9}) is part of the
$\mbox{USp}(2N)$ irreducibility conditions whereas (\ref{6.10})
relates the conformal dimension to the sum of the Dynkin labels:
\begin{equation}\label{6.12}
  \ell = 2\sum_{k=1}^N a_k\;.
\end{equation}
Let us denote the highest-weight UIR's of the $\mbox{OSp}(8^*/2N)$
algebra by
$$
{\cal D}(\ell;J_1,J_2,J_3;a_1,\ldots,a_N)
$$
where $\ell$ is the conformal dimension, $J_1,J_2,J_3$ are the
$\mbox{SU}^*(4)$ Dynkin labels and $a_k$ are the $\mbox{USp}(2N)$
Dynkin labels of the first component. Then the G-analytic
superfields defined above are of the type
\begin{equation}\label{6.13}
 \Phi(\theta^{1,2,\ldots,2N-1}) \ \Leftrightarrow \
{\cal D}(2\sum_{k=1}^N a_k;0,0,0;a_1,\ldots,a_N)\;.
\end{equation}

From the superconformal algebra it is clear that we can go on in
the same manner until we remove half of the $\theta$'s, namely
$\theta^{N+1},\ldots,\theta^{2N}$. Each time we have to set a new
Dynkin label to zero. We can summarize by saying that the
superconformal algebra $\mbox{OSp}(8^*/2N)$ admits the following
short UIR's corresponding to BPS states:
\begin{equation}\label{6.18}
  {p\over 2N}\mbox{ BPS}:\quad
{\cal D}(2\sum_{k=p}^N a_k;0,0,0;0,\ldots,0,a_p,\ldots,a_N)\;,
\quad p=1,\ldots,N\;.
\end{equation}

\subsection{Supersingletons}

There exist three types of massless multiplets in six dimensions
corresponding to ultrashort UIR's (supersingletons) of
$\mbox{OSp}(8^*/2N)$ (see, e.g., \cite{GT} for the case $N=2$).
All of them can be formulated in terms of constrained superfields
as follows.

{\sl (i)} The first type is described by a superfield
$W^{\{i_1\ldots i_n\}}(x,\theta)$, $1\leq n \leq N$, which is
antisymmetric and traceless in the external $\mbox{USp}(2N)$
indices (for even $n$ one can impose a reality condition). It
satisfies the constraint (see \cite{HSiT} and \cite{Park})
\begin{equation}\label{6.19}
  D^{(k}_\alpha W^{\{i_1)i_2\ldots i_n\}}=0 \qquad \Rightarrow \ {\cal
D}(2;0,0,0;0,\ldots,0,a_{n}=1,0,\ldots,0)
\end{equation}
The components of this superfield are massless fields. In the case
$N=n=1$ this is the on-shell $(1,0)$ hypermultiplet and for
$N=n=2$ it is the on-shell $(2,0)$ tensor multiplet
\cite{HSiT,bsvp}.

{\sl (ii)} The second type is described by a (real) superfield
without external indices, $w(x,\theta)$ obeying the constraint
\begin{equation}\label{6.23}
 D^{(i}_{[\alpha} D^{j)}_{\beta]} w = 0 \qquad \Rightarrow \ {\cal
D}(2;0,0,0;0,\ldots,0)\;.
\end{equation}

{\sl (iii)} Finally, there exists an infinite series of
multiplets described by superfields with $n$ totally symmetrized
external Lorentz spinor indices,
$w_{(\alpha_1\ldots\alpha_n)}(x,\theta)$ (they can be made real
in the case of even $n$) obeying the constraint
\begin{equation}\label{6.24}
  D^i_{[\beta}w_{(\alpha_1]\ldots\alpha_n)} = 0 \qquad \Rightarrow \ {\cal
D}(2+n/2;n,0,0;0,\ldots,0)\;.
\end{equation}

As shown in ref. \cite{FS3}, the six-dimensional massless
conformal fields only carry reps $(J_1,0)$ of the little group
$\mbox{SU}(2)\times \mbox{SU}(2)$ of a light-like particle
momentum. This result is related to the analysis of conformal
fields in $d$ dimensions \cite{Siegel1,AL}. This fact implies
that massless superconformal multiplets are classified by a
single $\mbox{SU}(2)$ and $\mbox{USp}(2N)$ R-symmetry and are
therefore identical to massless super-Poincar\'e multiplets in
five dimensions. Some physical implication of the above
circumstance have recently been discussed in ref. \cite{HULL2}
where it was suggested that certain strongly coupled $d=5$
theories effectively become six-dimensional.

\subsection{Harmonic superspace}

The massless multiplets {\sl (i), (ii)} admit an alternative
formulation in harmonic superspace (see \cite{HStT,Zp,Howe} for
$N=1,2$). The advantage of this formulation is that the
constraints (\ref{6.19}) become conditions for G-analyticity. We
introduce harmonic variables describing the coset
$\mbox{USp}(2N)/[\mbox{U}(1)]^N$:
\begin{equation}\label{6.25}
  u\in \mbox{USp}(2N): \qquad u^I_iu^i_J = \delta^I_J\;,
\ \ u^I_i \Omega^{ij}u^J_j = \Omega^{IJ}\;, \ \  u^I_i=
(u^i_I)^*\;.
\end{equation}
Here the indices $i,j$ belong to the fundamental representation of
$\mbox{USp}(2N)$ and $I,J$ are labels corresponding to the
$[\mbox{U}(1)]^N$ projections. The harmonic derivatives
\begin{equation}\label{6.26}
  D^{IJ} = \Omega^{K(I}u^{J)}_i{\partial\over\partial u^K_i}
\end{equation}
form the algebra of $\mbox{USp}(2N)_R$ realized on the indices
$I,J$ of the harmonics.

Let us now project the defining constraint (\ref{6.19}) with the
harmonics $u^K_k u^1_{i_1}\ldots u^n_{i_n}$, $K=1,\ldots,n$:
\begin{equation}\label{6.27}
D^1_\alpha W^{12\ldots n} = D^2_\alpha W^{12\ldots n}= \ldots =
D^n_\alpha W^{12\ldots n} =0
\end{equation}
where $D^{K}_\alpha = D^i_\alpha u^{K}_i$ and $W^{12\ldots
n}=W^{\{i_1\ldots i_n\}}u^1_{i_1}\ldots u^n_{i_n}$. Indeed, the
constraint (\ref{6.19}) now takes the form of a G-analyticity
condition. In the appropriate basis in superspace the solution to
(\ref{6.27}) is a short superfield depending on part of the odd
coordinates:
\begin{equation}\label{6.28}
W^{12\ldots n}(x_A,\theta^1,\theta^2,\ldots, \theta^{2N-n},u)\;.
\end{equation}
In addition to (\ref{6.27}), the projected superfield $W^{12\ldots
n}$ automatically satisfies the $\mbox{USp}(2N)$ harmonic
irreducibility conditions
\begin{equation}\label{6.29}
   D^{K\; 2N-K}W^{12} = 0\;, \quad K=1,\ldots,N
\end{equation}
(only the simple roots of $\mbox{USp}(2N)$ are shown). The
equivalence between the two forms of the constraint follows from
the obvious properties of the harmonic products $u^K_{[k} u^K_{i]}
=0$ and $\Omega^{ij}u^K_iu^L_j=0$ for $1\leq K < L\leq n$. The
harmonic constraints (\ref{6.29}) make the superfield ultrashort.

Finally, in case (ii), projecting the constraint (\ref{6.23}) with
$u^I_iu^I_j$ where $I=1,\ldots,N$ (no summation), we obtain the
condition
\begin{equation}\label{6.30'}
  D^I_\alpha D^I_\beta w=0\;.
\end{equation}
It implies that the superfield $w$ is {\sl linear} in each
projection $\theta^{\alpha I}$.

\subsection{Series of UIR's of $\mbox{OSp}(8^*/2N)$ and shortening}
\label{short6}

It is now clear that we can realize the BPS series of UIR's
(\ref{6.18}) as products of the different G-analytic superfields
(supersingletons) (\ref{6.27}).\fnm{a}\fnt{a}{As a bonus, we also
prove the unitarity of these series, since they are obtained by
multiplying massless unitary multiplets.} BPS shortening is
obtained by setting the first $p-1$ $\mbox{USp}(2N)$ Dynkin
labels to zero:
\begin{equation}\label{6.34}
{p\over 2N}\ \mbox{BPS}:\ \ W^{[0,\ldots,0,a_p,\ldots,a_N]}
(\theta^1,\theta^2,\ldots,\theta^{2N-p}) =  (W^{1\ldots
p})^{a_p}\ldots (W^{1\ldots N})^{a_N}
\end{equation}
(note that even if $a_1\neq 0$ we still have $1/2N$ shortening).

We remark that our harmonic coset $\mbox{USp}(2N)/[\mbox{U}(1)]^N$
is effectively reduced to
\begin{equation}\label{6.35}
  {\mbox{USp}(2N)\over \mbox{U}(p)\times [\mbox{U}(1)]^{N-p}}
\end{equation}
in the case of $p/2N$ BPS shortening (just as it happened in four
dimensions).  Such a smaller harmonic space was used in  Ref.
\cite{Howe} to formulate the $(2,0)$ tensor multiplet.

A study of the most general UIR's of $\mbox{OSp}(8^*/2N)$
(similar to the one of Ref. \cite{dp} for the case of
$\mbox{SU}(2,2/N)$) is presented in Ref. \cite{Minw2}. We can
construct these UIR's by multiplying the three types of
supersingletons above:
\begin{equation}\label{6.32}
  w_{\alpha_1\ldots\alpha_{m_1}}w_{\beta_1\ldots\beta_{m_2}}
w_{\gamma_1\ldots\gamma_{m_3}}\; w^k\; W^{[a_1,\ldots,a_N]}
\end{equation}
where $m_1\geq m_2 \geq m_3$ and the spinor indices are arranged
so that they form an $\mbox{SU}^*(4)$ UIR with Young tableau
$(m_1,m_2,m_3)$ or Dynkin labels
$J_1=m_1-m_2,J_2=m_2-m_3,J_3=m_3$. Thus we obtain four distinct
series:
\begin{eqnarray}
  \mbox{A)}&& \ell
\geq 6 +{1\over 2}(J_1+2J_2+3J_3)+2\sum_{k=1}^N a_k\;; \nonumber\\
  \mbox{B)}&& J_3=0\;,  \qquad \ell
\geq 4 +{1\over 2}(J_1+2J_2)+2\sum_{k=1}^N a_k\;; \nonumber\\
  \mbox{C)}&& J_2=J_3=0\;, \qquad \ell
\geq 2 +{1\over 2}J_1+2\sum_{k=1}^N a_k\;; \nonumber\\
  \mbox{D)}&& J_1=J_2=J_3= 0\;, \qquad \ell
= 2\sum_{k=1}^N a_k\;. \label{6.33'}
\end{eqnarray}
The superconformal bound is saturated when $k=0$ in (\ref{6.32}).
Note that the values of the conformal dimension we can obtain are
``quantized" since the factor $w^k$ has $\ell=2k$ and $k$ must be
a non-negative integer to ensure unitarity. With this restriction
eq. (\ref{6.33'}) reproduces the results of Ref. \cite{Minw2}.
However, we cannot comment on the existence of a ``window" of
dimensions $2 +{1\over 2}J_1+2\sum_{k=1}^N a_k\leq \ell \leq 4
+{1\over 2}J_1+2\sum_{k=1}^N a_k$ conjectured in \cite{Minw2}.
\fnm{b}\fnt{b}{In a recent paper \cite{FFr} the UIR's of the
six-dimensional conformal algebra $\mbox{SO}(2,6)$ have been
classified. Note that the superconformal bound in case A (with
all $a_i=0$) is stronger that the purely conformal unitarity
bounds found in \cite{FFr}.}

In the generic case the multiplet (\ref{6.32}) is ``long", but for
certain special values of the dimension some shortening can take
place \cite{Minw2}.

\setcounter{equation}0
\section{The three-dimensional case}

In this section we carry out the analysis of the $d=3$ $N=8$
superconformal algebra $\mbox{OSp}(8/4,\mathbb{R})$ in a way
similar to the above. Some of the results have already been
presented in \cite{FS2}. As in the previous cases, our results
could easily be extended to $\mbox{OSp}(N/4,\mathbb{R})$
superalgebras with arbitrary $N$. The $N=2$ and $N=3$ cases were
considered in Ref. \cite{Torino}.

\subsection{The conformal superalgebra $\mbox{OSp}(8/4,\mathbb{R})$
and Grassmann analyticity}\label{CSGA}

The standard realization of the conformal superalgebra
$\mbox{OSp}(8/4,\mathbb{R})$ makes use of the superspace
\begin{equation}
{\mathbb R}^{3\vert 16} = {\mbox{OSp}(8/4,\mathbb{R})\over
\{K,S,M,D,T\}} = (x^\mu,\theta^{\alpha\; i})\;. \label{7.5}
\end{equation}
In order to study G-analyticity we need to decompose the
generators $Q^i_\alpha$ under $[\mbox{U}(1)]^4\subset
\mbox{SO}(8)$. Besides the vector representation $8_v$ of
$\mbox{SO}(8)$ we are also going to use the spinor ones, $8_s$ and
$8_c$. In this context we find it convenient to introduce the four
subgroups $\mbox{U}(1)$  by successive reductions: $\mbox{SO}(8)\
\rightarrow \ \mbox{SO}(2)\times \mbox{SO}(6)\sim
\mbox{U}(1)\times \mbox{SU}(4) \ \rightarrow \
[\mbox{SO}(2)]^2\times \mbox{SO}(4)\sim [\mbox{U}(1)]^2\times
\mbox{SU}(2)\times \mbox{SU}(2) \ \rightarrow \ [\mbox{SO}(2)]^4
\sim [\mbox{U}(1)]^4$. Denoting the four $\mbox{U}(1)$ charges by
$\pm$, $(\pm)$, $[\pm]$ and $\{\pm\}$, we decompose the three
8-dimensional representations as follows:
\begin{eqnarray}
8_v:\quad Q^i &\rightarrow& Q^{\pm\pm}, \ Q^{(\pm\pm)}, \
Q^{[\pm]\{\pm\}},\label{7.10}\\
 8_s:\quad \phi^a &\rightarrow&
\phi^{+(+)[\pm]}, \ \phi^{-(-)[\pm]}, \ \phi^{+(-)\{\pm\}}, \
\phi^{-(+)\{\pm\}}\label{7.11}\\
 8_c:\quad  \sigma^{\dot a} &\rightarrow& \sigma^{+(+)\{\pm\}},
\ \sigma^{-(-)\{\pm\}}, \ \sigma^{+(-)[\pm]}, \
\sigma^{-(+)[\pm]}\label{7.12}
\end{eqnarray}

Let us denote a quasi primary superconformal field of the
$\mbox{OSp}(8/4,\mathbb{R})$ algebra by the quantum numbers of its
HWS:
\begin{equation}\label{555}
 {\cal D}(\ell; J; a_1,a_2,a_3,a_4)
\end{equation}
where $\ell$ is the conformal dimension, $J$ is the Lorentz spin
and $a_i$ are the Dynkin labels (see, e.g., \cite{FSS}) of the
$\mbox{SO}(8)$ R symmetry.

In order to build G-analytic superspaces we have to add one or
more projections of $Q^i_\alpha$ to the coset denominator. In
choosing the subset of projections we have to make sure that: i)
they anticommute among themselves; ii) the subset is closed under
the action of the raising operators of $\mbox{SO}(8)$. Then we
have to examine the consistency of the vanishing of the chosen
projections with the conformal superalgebra. Thus we find the
following sequence of G-analytic superspaces corresponding to BPS
states:
\begin{eqnarray}
 {1\over 8}  \mbox{ BPS}: && \left\{
  \begin{array}{l}
    q_\alpha^{++}\Phi=0\ \rightarrow \\
    \Phi(\theta^{++},\theta^{(\pm\pm)},\theta^{[\pm]\{\pm\}})\\
    {\cal D}(a_1+a_2 + {1\over 2}(a_3+a_4);0;a_1,a_2,a_3,a_4)
     \end{array}
 \right.\label{7.16}\\
 {1\over 4}  \mbox{ BPS}: &&\left\{
  \begin{array}{l}
    q_\alpha^{++}\Phi=q_\alpha^{(++)}\Phi=0\ \rightarrow \\
    \Phi(\theta^{++},\theta^{(++)},\theta^{[\pm]\{\pm\}})\\
    {\cal D}(a_2 + {1\over 2}(a_3+a_4);0;0,a_2,a_3,a_4)
      \end{array}
 \right. \label{7.17}\\
{3\over 8}  \mbox{ BPS}: &&\left\{
  \begin{array}{l}
    q_\alpha^{++}\Phi=q_\alpha^{(++)}\Phi=q_\alpha^{[+]\{+\}}\Phi=0\ \rightarrow \\
    \Phi(\theta^{++},\theta^{(++)},\theta^{[+]\{\pm\}},\theta^{[-]\{+\}})\\
    {\cal D}({1\over 2}(a_3+a_4);0;0,0,a_3,a_4)
  \end{array}
 \right. \label{7.18}\\
{1\over 2}  \mbox{ BPS (type I)}: &&\left\{
  \begin{array}{l}
    q_\alpha^{++}\Phi=q_\alpha^{(++)}\Phi=q_\alpha^{[+]\{\pm\}}\Phi=0\ \rightarrow \\
    \Phi(\theta^{++},\theta^{(++)},\theta^{[+]\{\pm\}})\\
    {\cal D}({1\over 2}a_3;0;0,0,a_3,0)
  \end{array}
 \right. \label{7.19}\\
{1\over 2}  \mbox{ BPS (type II)}: &&\left\{
  \begin{array}{l}
    q_\alpha^{++}\Phi=q_\alpha^{(++)}\Phi=q_\alpha^{[\pm]\{+\}}\Phi=0\ \rightarrow \\
    \Phi(\theta^{++},\theta^{(++)},\theta^{[\pm]\{+\}})\\
    {\cal D}({1\over 2}a_4;0;0,0,0,a_4)
  \end{array}
 \right. \label{7.20}
\end{eqnarray}
Note the existence of two types of $1/2$ BPS states due to the two
possible subsets of projections of $q^i$ closed under the raising
operators of $\mbox{SO}(8)$.

\subsection{Supersingletons and harmonic superspace}

The supersingletons are the simplest $\mbox{OSp}(8/4,\mathbb{R})$
representations of the type (\ref{7.19}) or (\ref{7.20}) and
correspond to ${\cal D}(1/2; 0; 0,0,1,0)$ or ${\cal D}(1/2; 0;
0,0,0,1)$. The existence of two distinct types of $d=3$ $N=8$
supersingletons has first been noted in Ref. \cite{GNST}. Each of
them is just a collection of eight Dirac supermultiplets \cite{Fr}
made out of ``Di" and ``Rac" singletons \cite{ff2}.

In order to realize the supersingletons in superspace we note
that the HWS in the two supermultiplets above has spin 0 and the
Dynkin labels of the $8_s$ or $8_c$ of $\mbox{SO}(8)$,
correspondingly. Therefore we take a scalar superfield
$\Phi_a(x^\mu, \theta^\alpha_i)$ (or $\Sigma_{\dot a}(x^\mu,
\theta^\alpha_i)$) carrying an external $8_s$ index $a$ (or an
$8_c$ index $\dot a$). These superfields are subject to the
following on-shell constraints \fnm{c}\fnt{c}{See also \cite{Howe}
for the description of a supersingleton related to ours by
$\mbox{SO}(8)$ triality. Superfield representations of other
$OSp(N/4)$ superalgebras have been considered in \cite{IS,FFre}.}:
\begin{eqnarray}
  \mbox{type I:}&&D^i_\alpha\Phi_a = {1\over 8}\gamma^i_{a\dot
b}\tilde\gamma^j_{\dot b c} D^j_\alpha\Phi_c\;; \label{7.25}\\
  \mbox{type II:}&& D^i_\alpha\Sigma_{\dot a} = {1\over
8}\tilde\gamma^i_{\dot a b}\gamma^j_{b\dot c}
D^j_\alpha\Sigma_{\dot c}\;. \label{7.26}
\end{eqnarray}
The two multiplets consist of a massless scalar in the $8_s$
($8_c$) and spinor in the $8_c$ ($8_s$).

The harmonic superspace description of these supersingletons can
be realized by taking the harmonic coset \fnm{d}\fnt{d}{A
formulation of the above multiplet in harmonic superspace has
been proposed in Ref. \cite{Howe} (see also \cite{ZK} and
\cite{HL} for a general discussion of three-dimensional harmonic
superspaces). The harmonic coset used in \cite{Howe} is
$\mbox{Spin}(8)/\mbox{U}(4)$. Although the supersingleton itself
does indeed live in this smaller coset (see Section \ref{7.4.4}),
its residual symmetry $U(4)$ would not allow us to multiply
different realizations of the supersingleton. For this reason we
prefer from the very beginning to use the coset (\ref{7.27'})
with a minimal residual symmetry.}
\begin{equation}\label{7.27'}
  {\mbox{SO}(8)\over [\mbox{SO}(2)]^4} \ \sim \ {\mbox{Spin}(8)\over
 [\mbox{U}(1)]^4}\;.
\end{equation}
Since $\mbox{SO}(8)\sim \mbox{Spin}(8)$  has three inequivalent
fundamental representations, $8_s,8_c,8_v$, following  \cite{GHS}
we introduce three sets of harmonic variables:
\begin{equation}\label{7.27}
  u_a^A\;, \ w^{\dot A}_{\dot a}\;, \ v^I_i
\end{equation}
where $A$, $\dot A$ and $I$ denote the decompositions of an $8_s$,
$8_c$ and $8_v$ index, correspondingly, into sets of four
$\mbox{U}(1)$ charges (see (\ref{7.10})-(\ref{7.12})). Each of the
$8\times 8$ real matrices (\ref{7.27}) belongs to the
corresponding representation of $\mbox{SO}(8)\sim \mbox{Spin}(8)$.
This implies that they are orthogonal matrices (this is a
peculiarity of $\mbox{SO}(8)$ due to triality):
\begin{equation}\label{7.28}
  u_a^A u_a^B = \delta^{AB}\;, \quad w^{\dot A}_{\dot a} w^{\dot B}_{\dot
a}  = \delta^{\dot A\dot B}\;, \quad v^I_i v^J_i = \delta^{IJ} \;.
\end{equation}

Further, we introduce harmonic derivatives (the covariant
derivatives on the coset (\ref{7.27'})):
\begin{equation}\label{7.30}
  D^{IJ} = u^A_a (\gamma^{IJ})^{AB}{\partial\over\partial u^B_a} +
w^{\dot A}_{\dot a} (\gamma^{IJ})^{\dot A\dot
B}{\partial\over\partial w^{\dot B}_{\dot a}} + v^{[I}_i
{\partial\over\partial v^{J]}_{i}}\;.
\end{equation}
They respect the algebraic relations  (\ref{7.28}) among the
harmonic variables and form the algebra of $\mbox{SO}(8)$
realized on the indices $A,\dot A, I$ of the harmonics.

We now use the harmonic variables for projecting the
supersingleton defining constraints (\ref{7.25}), (\ref{7.26}). It
is easy to show that the projections $\Phi^{+(+)[+]}$ and
$\Sigma^{+(+)\{+\}}$ satisfy the following G-analyticity
constraints:
\begin{eqnarray}
  &&D^{++}\Phi^{+(+)[+]} = D^{(++)}\Phi^{+(+)[+]}=D^{[+]\{\pm\}}
\Phi^{+(+)[+]} = 0\;, \label{7.31}\\
  &&D^{++}\Sigma^{+(+)\{+\}} = D^{(++)}\Sigma^{+(+)\{+\}}=D^{[+]\{\pm\}}
\Sigma^{+(+)\{+\}} = 0 \label{7.31'}
\end{eqnarray}
where $D^I_\alpha = v^I_iD^i_\alpha$, $\Phi^A = u^A_a\Phi_a$ and
$\Sigma^{\dot A} = w^{\dot A}_{\dot a}\Sigma_{\dot a}$. This is
the superspace realization of the 1/2 BPS shortening conditions
(\ref{7.19}), (\ref{7.20}). In the appropriate basis in superspace
$\Phi^{+(+)[+]}$ and $\Sigma^{+(+)\{+\}}$ depend on different
halves of the odd variables as well as on the harmonic variables:
\begin{eqnarray}
  \mbox{type I}:&& \Phi^{+(+)[+]}
(x_A,\theta^{++}, \theta^{(++)}, \theta^{[+]\{\pm\}}, u,w) \;,
\label{7.32}\\
 \mbox{type II}: && \Sigma^{+(+)\{+\}}(x_A,\theta^{++},\theta^{(++)},
\theta^{[\pm]\{+\}}, u,w)\;.\label{7.32'}
\end{eqnarray}

In addition to the G-analyticity constraints (\ref{7.31}),
(\ref{7.31'}), the on-shell superfields $\Phi^{+(+)[+]}$,
$\Sigma^{+(+)\{+\}}$ are subject to the $\mbox{SO}(8)$
irreducibility harmonic conditions obtained by replacing the
$\mbox{SO}(8)$ raising operators by the corresponding harmonic
derivatives. The combination of the latter with eq. (\ref{7.31})
is equivalent to the original constraint (\ref{7.25}).

\subsection{$\mbox{OSp}(8/4,\mathbb{R})$ supersingleton composites}

One way to obtain short multiplets of $\mbox{OSp}(8/4,\mathbb{R})$
is to multiply different analytic superfields describing the type
I supersingleton. The point is that above we chose a particular
projection of, e.g., the defining constraint (\ref{7.25}) which
lead to the analytic superfield  $\Phi^{+(+)[+]}$. In fact, we
could have done this in a variety of ways, each time obtaining
superfields depending on different halves of the total number of
odd variables. Leaving out the $8_v$ lowest weight $\theta^{--}$,
we can have four distinct but equivalent analytic descriptions of
the type I supersingleton:
\begin{eqnarray}
  &&\Phi^{+(+)[+]}
(\theta^{++}, \theta^{(++)}, \theta^{[+]\{+\}},
\theta^{[+]\{-\}})\;, \nonumber\\
  &&\Phi^{+(+)[-]} (\theta^{++}, \theta^{(++)},
\theta^{[-]\{+\}}, \theta^{[-]\{-\}})\;, \nonumber\\
  &&\Phi^{+(-)\{+\}} (\theta^{++},
\theta^{(--)}, \theta^{[+]\{+\}}, \theta^{[-]\{+\}})\;,
\nonumber\\
  &&\Phi^{+(-)\{-\}}
(\theta^{++}, \theta^{(--)}, \theta^{[+]\{-\}},
\theta^{[-]\{-\}})\;. \label{7.34}
\end{eqnarray}
Then we can multiply them in the following way:
\begin{equation}\label{7.35}
  (\Phi^{+(+)[+]})^{p+q+r+s}(\Phi^{+(+)[-]})^{q+r+s}
(\Phi^{+(-)\{+\}})^{r+s}(\Phi^{+(-)\{-\}})^{s}
\end{equation}
thus obtaining three series of $\mbox{OSp}(8/4,\mathbb{R})$ UIR's
exhibiting $1/8$, $1/4$ or $1/2$ BPS shortening:
\begin{eqnarray}
 {1\over 8}  \mbox{ BPS:} && {\cal D}(a_1+a_2 + {1\over
2}(a_3+a_4), 0; a_1,a_2,a_3,a_4)\;, \quad a_1-a_4 = 2s \geq 0\;;
 \nonumber\\
 {1\over 4}  \mbox{ BPS:} && {\cal D}(a_2 + {1\over 2}a_3, 0;
0,a_2,a_3,0)\;;  \label{7.36}\\
 {1\over 2}  \mbox{ BPS:} && {\cal D}({1\over 2}a_3, 0; 0,0,a_3,0)
\nonumber \end{eqnarray}
where
\begin{equation}\label{7.37'}
a_1=r+2s\;, \quad a_2= q\;, \quad a_3=p\;, \quad a_4=r\;.
\end{equation}

We see that multiplying only one type of supersingletons cannot
reproduce the general result of Section \ref{CSGA} for all
possible short multiplets. Most notably, in (\ref{7.36}) there is
no $3/8$ series. The latter can be obtained  by mixing the two
types of supersingletons:
\begin{equation}\label{7.37}
[\Phi^{+(+)[+]}(\theta^{++},\theta^{(++)},\theta^{[+]\{\pm\}})]^{a_3}
[\Sigma^{+(+)\{+\}}(\theta^{++},\theta^{(++)},\theta^{[\pm]\{+\}})]^{a_4}
\end{equation}
(or the same with $\Phi$ and $\Sigma$ exchanged). Counting the
charges and the dimension, we find exact matching with the series
(\ref{7.18}):
\begin{equation}\label{7.37''}
 {3\over 8}  \mbox{ BPS:} \quad {\cal D}({1\over 2}(a_3+a_4);0;0,0,a_3,a_4)\;.
\end{equation}
Further, mixing two realizations of type I and one of type II
supersingletons, we can construct the 1/4 series
\begin{equation}\label{7.38}
   [\Phi^{+(+)[+]}]^{a_2+a_3}[\Phi^{+(+)[-]}]^{a_2}
[\Sigma^{+(+)\{+\}}]^{a_4}
\end{equation}
which corresponds to (\ref{7.17}):
\begin{equation}\label{7.37'''}
 {1\over 4}  \mbox{ BPS:} \quad
{\cal D}(a_2 + {1\over 2}(a_3+a_4);0;0,a_2,a_3,a_4)\;.
\end{equation} Finally, the full 1/8 series
(\ref{7.16}) (i.e., without the restriction $a_1-a_4 = 2s\geq 0$
in (\ref{7.36})) can be obtained in a variety of ways.

\subsection{BPS states of $\mbox{OSp}(8/4,\mathbb{R})$}

Here we give a summary of all possible
$\mbox{OSp}(8/4,\mathbb{R})$ BPS multiplets. Denoting the UIR's by
\begin{equation}\label{7.39}
  {\cal D}(\ell;J;a_1,a_2,a_3,a_4)
\end{equation}
where $\ell$ is the conformal dimension, $J$ is the spin and
$a_1,a_2,a_3,a_4$ are the $\mbox{SO}(8)$ Dynkin labels, we find
four BPS conditions:

\subsubsection{}

\begin{equation}\label{8.14}
  {1\over 8}\ \mbox{BPS}:\qquad  q_\alpha^{++} =0\;.
\end{equation}
The corresponding UIR's are:
\begin{equation}\label{8.15}
  {\cal D}(a_1+a_2 + {1\over 2}(a_3+a_4);0;a_1,a_2,a_3,a_4)
\end{equation}
and the harmonic coset is
\begin{equation}\label{8.16}
  {\mbox{Spin}(8)\over [\mbox{U}(1)]^4}\;.
\end{equation}
If $a_2=a_3=a_4=0$ this coset becomes
$\mbox{Spin}(8)/\mbox{U}(4)$.

\subsubsection{}

\begin{equation}\label{8.17}
  {1\over 4}\ \mbox{BPS}:\qquad  q_\alpha^{++} =q_\alpha^{(++)} =0\;.
\end{equation}
The corresponding UIR's are:
\begin{equation}\label{8.18}
  {\cal D}(a_2 + {1\over 2}(a_3+a_4);0;0,a_2,a_3,a_4)
\end{equation}
and the harmonic coset is
\begin{equation}\label{8.19}
  {\mbox{Spin}(8)\over [\mbox{U}(1)]^2\times \mbox{U}(2)}\;.
\end{equation}
If $a_3=a_4=0$ this coset becomes
$\mbox{Spin}(8)/\mbox{U}(1)\times [\mbox{SU}(2)]^3$.

\subsubsection{}

\begin{equation}\label{8.20}
  {3\over 8}\ \mbox{BPS}:\qquad  q_\alpha^{++} =q_\alpha^{(++)} =
q_\alpha^{[+]\{+\}} =0\;.
\end{equation}
The corresponding UIR's are:
\begin{equation}\label{8.21}
  {\cal D}({1\over 2}(a_3+a_4);0;0,0,a_3,a_4)
\end{equation}
and the harmonic coset is
\begin{equation}\label{8.22}
  {\mbox{Spin}(8)\over \mbox{U}(1)\times \mbox{U}(3)}\;.
\end{equation}

\subsubsection{} \label{7.4.4}

\begin{eqnarray}
  {1\over 2}\ \mbox{BPS (type I)}:&& q_\alpha^{++} =q_\alpha^{(++)} =
q_\alpha^{[+]\{+\}}=q_\alpha^{[+]\{\pm\}} =0\;; \label{8.23}\\
  {1\over 2}\ \mbox{BPS (type II)}:&& q_\alpha^{++} =q_\alpha^{(++)} =
q_\alpha^{[+]\{+\}}=q_\alpha^{[\pm]\{+\}} =0\;. \label{8.23'}
\end{eqnarray}
The corresponding UIR's are:
\begin{eqnarray}
  {1\over 2}\ \mbox{BPS (type I)}:&& {\cal D}({1\over 2}a_3;0;0,0,a_3,0)\;; \label{8.24}\\
  {1\over 2}\ \mbox{BPS (type II)}:&& {\cal D}({1\over 2}a_4;0;0,0,0,a_4)\;. \label{8.24'}
\end{eqnarray}
and the harmonic coset is
\begin{equation}\label{8.25}
  {\mbox{Spin}(8)\over \mbox{U}(4)}\;.
\end{equation}

\pagebreak

\nonumsection{Acknowledgements} \noindent

E.S. is grateful to the TH Division of CERN for its kind
hospitality. The work of S.F. has been supported in part by the
European Commission TMR programme ERBFMRX-CT96-0045 (Laboratori
Nazionali di Frascati, INFN) and by DOE grant DE-FG03-91ER40662,
Task C.


\nonumsection{References}

\end{document}